\documentclass[twocolumn,10pt,final]{adf_elsart}
\usepackage{verbatim}
\usepackage{graphicx,floatfig}
\usepackage{float}
\usepackage[latin1]{inputenc}
\usepackage{t1enc}
\usepackage{times}
\usepackage{mathptm}
\usepackage{color}

\let \IG \includegraphics
\begin{document}
\initfloatingfigs

\begin{frontmatter}
\title{Demonstration of detuned dual recycling at the Garching 30\,m laser interferometer}

\author[mpq]{A. Freise\thanksref{email_adf}}
\author[mpq]{G. Heinzel\thanksref{now_ghh}}
\author[glasgow]{K.A. Strain}
\author[mpq]{J. Mizuno\thanksref{now_jum}}
\author[glasgow]{K.D. Skeldon}
\author[amp]{H. L\"uck}
\author[amp]{B. Willke}
\author[mpq]{R. Schilling}
\author[mpq]{A. R\"udiger}
\author[mpq]{W. Winkler}
\author[amp,mpq]{K. Danzmann}
\address[mpq]{Max-Planck-Institut f\"ur Quantenoptik, Hans-Kopfermann-Str. 1, D-85748 Garching, Germany}
\address[amp]{Institut f\"ur Atom- und Molek\"ulphysik, Universit\"at Hannover, Callinstr. 38, D-30167 Hannover, Germany}
\address[glasgow]{University of Glasgow, Department of Physics and Astronomy, Glasgow G12 8QQ, Great Britain}
\thanks[email_adf]{adf@mpq.mpg.de}
\thanks[now_ghh]{Now at: Space-Time Astronomy Section, National Astronomical Observatory, 2--21--1 Ohsawa, Mitaka, Tokyo 181--8588, Japan}
\thanks[now_jum]{Now at: Photonic Technology Division, Communications Research Laboratory (CRL),4-2-1 Nukuikitamachi, Koganei 184-8795 JAPAN}

\begin{abstract} 
Dual recycling is an advanced optical technique to enhance the 
signal-to-noise ratio of laser interferometric gravitational wave
detectors in a limited bandwidth.
To optimise the center of this band with respect to Fourier
frequencies of expected gravitational wave signals,
{\it detuned} dual recycling
has to be implemented. We have demonstrated detuned dual recycling
on a fully suspended 30\,m prototype interferometer. A control scheme 
that allows
the detector to be tuned to different frequencies will be outlined. 
Good agreement between the experimental results and 
numerical simulations has been achieved.
\end{abstract}

\end{frontmatter}

\section{Introduction}
Several long-baseline laser interferometers are currently under construction 
worldwide, aimed at the direct detection of gravitational 
waves~\cite{LIGO:prop,VIRGO:prop,GEO600:prop,TAMA:prop,ACIGA:prop}. 
These instruments are basically 
Michelson interferometers optimised to measure
tiny differential phase modulations of the light in the two arms, which
would be the main effect of a passing gravitational wave. 
In order to reduce the effect of 
the photon shot-noise, which is one of the main noise sources in these
detectors, the established technique of power recycling
is used to maximise the light power circulating in the arms.
The sensitivity can be further improved by optimising the 
signal storage time, i.e.\ the average time for which the 
phase modulation sidebands induced by the gravitational wave
are stored in the interferometer.
For this purpose 
the advanced techniques of {\em signal recycling\/} and {\em resonant 
sideband extraction\/}
have been proposed and demonstrated on prototypes~\cite{ghh:dr,ghh:rse}.

One of the proposed features of signal recycling is the possibility
to {\em detune\/} the detector in a way that allows the frequency of the
sensitivity maximum to be set to an arbitrary non-zero value. 
This is useful in order to optimise the sensitivity for the detection of 
signals at a predetermined frequency.
To use a technique like
detuned signal recycling in a gravitational wave detector a control 
scheme has to be developed.
A method is required (combining optical and electronical techniques)
to produce error signals for the microscopic position 
(within a small fraction of a wavelength) of each 
mirror, which can be used in feedback loops to set and maintain 
the desired operating point of the interferometer. 
For best versatility the control
scheme should include the {\em broadband\/} state (see below) as well 
as the full
intended detuning range. It should provide an easy way to change the
detuning, so that an electronic servo can be used to adjust
the amount of detuning (and hence the frequency of the sensitivity maximum)
during operation of the detector.
This paper describes a control scheme that offers these advantages
for the {\em dual recycling\/} system which will be used in GEO\,600.
For a more general description of the 30\,m prototype with dual
recycling see \cite{ghh:dr} and \cite{ghh:phd}.

\begin{figure}[htb]
\IG [bb=0 0 270 270,scale=.7] {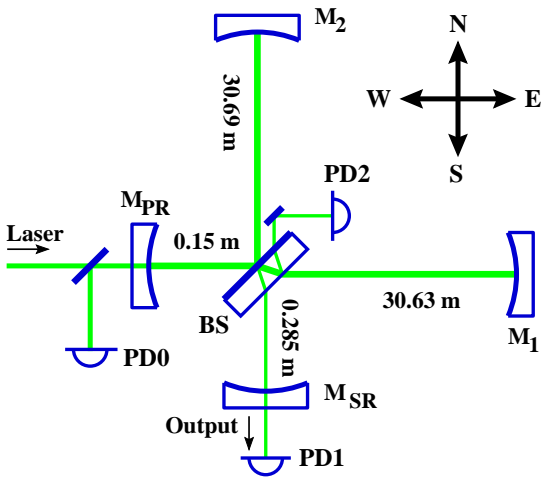} 
\caption{The optical layout of the dual recycled interferometer.}
\label{fig:MI1}
\end{figure}

\section{Recycling techniques}
The Michelson interferometer (consisting of $\rm BS$, $\rm M_1$ and  $\rm M_2$) is operated at the {\em dark fringe\/}
in which case the light of the carrier frequency is reflected back towards the 
West (we labelled the different ports of the Michelson
interferometer North, South, East, and West, see Fig.\ \ref{fig:MI1})
and ideally never reaches the South port. A passing gravitational 
wave (or a differential motion of the
end mirrors $\rm M_1$ and $\rm M_2$) will modulate the phase of the
light in the interferometer arms and thus create `signal sidebands', which
are not directed to the West but to the South port. The signal can 
then be detected in the light intensity in the South port. This is
the main interferometer output signal.

\begin{figure}[htb]
\IG [scale=.6] {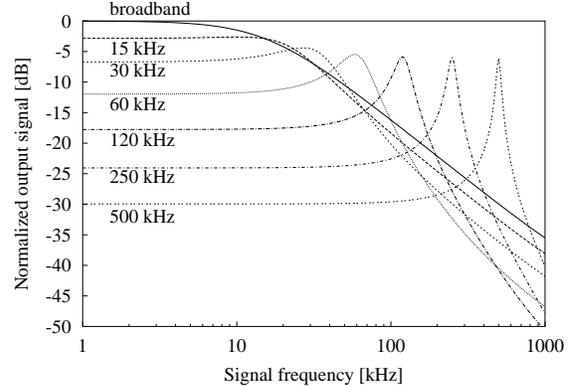} 
\caption{Simulation of the 30\,m prototype interferometer response 
tuned to different frequencies. The parameter that is changed to 
tune (or detune) the interferometer is the so-called 
`detuning' $\xi_{\rm SR}$, i.e. the microscopic position of the signal recycling mirror (Eq. \ref{eq:SR}).}
\label{fig:xfktns}
\end{figure}

Both signal recycling and resonant sideband extraction 
employ an additional mirror $\rm M_{\rm SR}$ at this
Michelson interferometer output port  
(see Fig.~\ref{fig:MI1}). For the signal sidebands  
a cavity (the {\em signal recycling 
cavity\/}) is formed by $\rm M_{\rm SR}$ 
and the Michelson interferometer. Note that in the dark fringe condition
the Michelson interferometer 
behaves like a mirror if seen from both the West and South ports.
In the nominal operating point the bandwidth
of the interferometer 
with signal recycling is determined by the
reflectivity of $\rm M_{\rm SR}$, when all losses inside the 
interferometer are smaller than the transmission of $\rm M_{\rm SR}$.
There are three possible modes of operation of the signal recycling
cavity, which 
are distinguished by the microscopic
position of $\rm M_{\rm SR}$. 
We describe that position by the `detuning' $\xi_{\rm SR}$,
defined by 
\begin{equation}
\xi_{\rm SR}=2\pi\,\frac{L_{\rm SR}}{\lambda_0} \,\,\,\,{\rm mod}\,\, \pi\label{eq:SR}
\end{equation}
where $L_{\rm SR}$
is the average distance between $\rm M_{\rm SR}$ and the two end mirrors
$\rm M_1$, $\rm M_2$ (see Eq. \ref{eq:LSR} and Fig.~\ref{fig:block1})
and $\lambda_0$ refers to
the carrier wavelength. The three modes of operation are:
\begin{enumerate}
\item{In {\em  broadband signal recycling}, $\rm M_{\rm SR}$ is tuned such that
the carrier frequency is resonant in the signal recycling cavity 
($\xi_{\rm SR}=0$). The sensitivity is enhanced for low 
Fourier frequencies, at the
expense of reducing the bandwidth to the 
range from DC to $\rm {\rm B_{\rm SR}} / 2$, where $\rm B_{\rm SR}$ is the FWHM
bandwidth of the signal recycling cavity.}
\item{In {\em resonant sideband extraction}, the signal recycling cavity is
tuned to {\em anti-resonance\/} for the carrier frequency 
($\xi_{\rm SR}=\pi/2$), which results in 
a wider bandwidth 
than that of the instrument without $\rm M_{\rm SR}$. This
scheme is useful only for detectors with Fabry-Perot cavities in the
arms and when the arm cavity bandwidth is made as narrow as technically
possible in order to maximise the energy stored in the arms.}
\item{Both signal recycling and resonant sideband extraction have, 
apart from their standard
mode of operation described above, a so-called {\em detuned\/} mode that has a 
sensitivity maximum at a selectable {\it non-zero} Fourier
frequency. 
Those modes are characterised by \mbox{$0 < \xi_{\rm SR} < \pi/2$}.
The width of the sensitivity peak is given by $\rm B_{\rm SR}$.
Detuned recycling thus creates the possibility of tuning the detector to the
exact frequency of an expected periodic signal or even of tracking
a changing periodic signal (chirp). 
Figure\ \ref{fig:xfktns} shows several computed transfer functions
(gravitational wave signal $\rightarrow$ detector output) 
for detuned and broadband  dual recycling.}
\end{enumerate}

Signal recycling in combination  with power recycling was proposed by 
Meers~\cite{meers:dr} and is called 
{\em dual 
recycling}. 
The first experimental demonstration was performed on a table-top setup  by
Strain et~al.~\cite{strain:dr}. 
Further table-top experiments have investigated various control 
schemes~\cite{gray:dr,Maass}. 
Dual recycling will be used from 
the beginning in
the British-German project GEO\,600 whilst also 
being considered for implementation
in the further stages of various other gravitational wave detectors.
Recently the first
experimental demonstration of dual recycling on a fully suspended 
interferometer at the 30\,m prototype in Garching has been 
reported~\cite{ghh:dr}. 
Subsequently, we have operated 
the 30\,m prototype in a {\em detuned dual recycling\/} 
mode. That will be 
the subject of this paper, and special emphasis will be given 
to the control scheme, which can be extended for use in
the gravitational wave detectors currently being built. 

Although some features of the
dual recycled interferometer can be predicted by simple models, the 
general properties of such a system are very
complex. Therefore a computer model has been 
developed~\cite{jun:optmat,adf:finesse}
which allows the computation of
transfer functions and error signals of a complex optical system like
a dual recycled interferometer. The interferometer
is simulated by solving a set of time independent linear equations that
represents the quasistatic light fields inside the interferometer
in the ray optics formalism. The simulation has been used to model
error signals 
and transfer functions and is also used to choose suitable operating points
of a dual recycled interferometer.

\section{Control signals}
\begin{figure}[bth]
\IG [scale=.55] {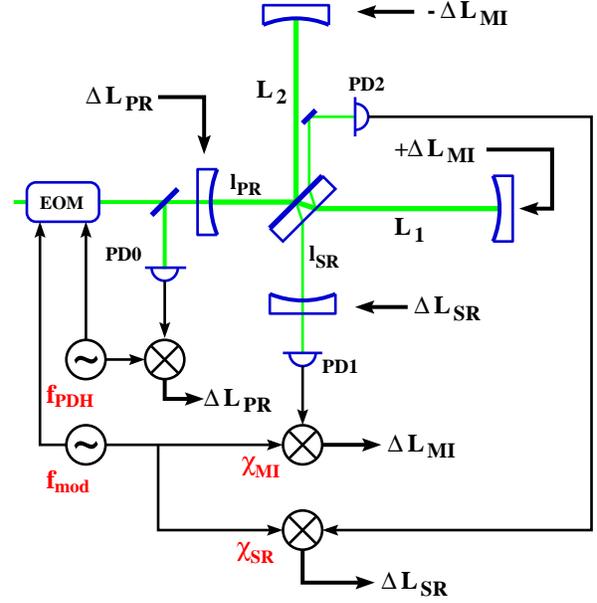} 
\caption{The error signals for the three control loops are generated 
by RF modulation techniques: two RF phase
modulations are applied to the light before it enters the interferometer,
and the photo currents detected at different positions of the
interferometer are demodulated with one of the RF frequencies respectively. 
The error signal for the length
of the power recycling cavity is detected in front of the interferometer
(PD0) using the frequency $f_{\rm PDH}$. The signal for the
differential motion
of the end mirrors $L_{\rm MI}$ is detected in the main output of the 
interferometer (PD1) using the frequency
$f_{\rm mod}$. Again $f_{\rm mod}$ is used to generate the error signal for the
signal recycling cavity length detected on PD2.}
\label{fig:block1}
\end{figure}

The dual recycled interferometer consists of 4 mirrors and the
beam splitter. The three degrees of freedom that have to be
controlled by electronic control systems to keep the interferometer at its
operating point are (see Fig.~\ref{fig:block1}):

\begin{enumerate}
\item{the length of the power recycling cavity 
\begin{equation}
L_{\rm PR}=l_{\rm PR}+(L_1+L_2)/2\label{eq:LPR}
\end{equation}}
\item{the length of the signal recycling cavity 
\begin{equation}
L_{\rm SR}=l_{\rm SR}+(L_1+L_2)/2\label{eq:LSR}
\end{equation}}
\item{the differential arm length of the Michelson interferometer 
\begin{equation}
L_{\rm MI}=L_1-L_2\label{eq:LMI}
\end{equation}}
\end{enumerate}

The length of the power recycling cavity can be controlled using a
standard Pound-Drever-Hall method~\cite{schnier:pr}. 
The control of the Michelson interferometer
and the signal recycling cavity length is more complicated and
depends on whether the system should be set to the {\em detuned\/}
mode or the {\em broadband} mode.
The control scheme described here
makes use of a modulation technique that was 
originally proposed by 
Schnupp~\cite{schnupp:schnupp}:
an RF phase modulation is applied to the light 
before it enters the interferometer,
and a macroscopic arm length difference causes some fraction of the 
modulation sidebands to appear at the dark fringe (South) port where
they act as local oscillator for the detection of the 
gravitational wave signal.
To distinguish these
RF phase modulation sidebands used for controlling the interferometer from
the sidebands induced by a gravitational wave,
the former will be called {\em Schnupp sidebands\/} in the following.

The state of the Michelson interferometer 
can be expressed by the 
differential microscopic detuning $\xi_{\rm MI}$ of the arm length 
difference, defined by:
\begin{equation}
\xi_{\rm MI}=2\pi\,\frac{L_{\rm MI}}{\lambda_0} \,\,\,\,{\rm mod}\,\, \pi
\end{equation}
In our case
$\xi_{\rm MI}=0$ corresponds to the dark fringe condition, the 
nominal operating point of all experiments discussed here.
Similarly, $\xi_{\rm SR}=0$ refers to the state of the signal recycling
cavity when the carrier light is on resonance
in this cavity ({\em broadband signal recycling mode\/}).

{\bf Broadband recycling} control signals can be understood with
an intuitive model in which the interferometer contains two coupled cavities
(the signal recycling and power recycling cavity) 
resonant for the carrier frequency.
In the dark fringe condition, 
ideally the carrier light only is present in the power recycling cavity and 
never reaches the signal recycling mirror, whereas the signal sidebands
are circulating only in the signal recycling cavity. With broadband
recycling the resonance of the 
signal recycling cavity is centered on the carrier
frequency,
and for signal frequencies within the SR cavity bandwidth both signal 
sidebands are resonantly enhanced.

Simulations show
that suitable error signals for the length control system of the Michelson
interferometer and the signal recycling cavity 
can be obtained from the light in the South port
and light split off from one of the interferometer arms respectively
(e.g.\ using the demodulated
photocurrent of PD1 and PD2 in Fig.\ \ref{fig:block1}). If the Michelson 
fluctuates around its 
operating point $\xi_{\rm MI}=0$, some carrier light leaves the 
interferometer at the South port. Beating this light 
with the Schnupp modulation sidebands
thus generates an error signal at PD1 after demodulation. Similarly, sidebands induced by a
gravitational wave are directed to the South port and generate a 
detectable signal. The South port is then
the main output of the interferometer, and PD1 provides the gravitational wave
signal as well as the error signal for the Michelson control.

When the tuning of the signal recycling mirror changes, 
the Schnupp modulation sidebands 
experience a phase shift, which can only be detected by beating with the
carrier light which is not present in the South port but in the rest
of the interferometer. This beat signal supplies  
an error signal that can
be detected either in the West, East, or North port 
(it was detected by PD2 in our experiment).

\begin{figure}[htb]
\centering
\IG [scale=.8] {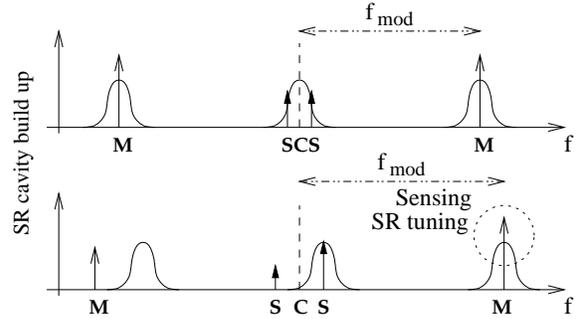} 
\caption{Resonance conditions in the signal recycling cavity for the broadband
mode (upper trace) and the detuned case (lower trace). M indicates the Schnupp
modulation sidebands, S the gravitational wave sidebands and C the carrier.}
\label{fig:det}
\end{figure}

The macroscopic 
arm length difference of the Michelson interferometer determines 
the amplitude of the Schnupp sidebands
transferred from the power recycling cavity into the signal recycling 
cavity. 
Ideally the shot noise limited sensitivity of the 
Michelson interferometer to gravitational
wave signals is independent of the amplitude of
the Schnupp sidebands reaching the South port. 
In real systems it is necessary that the modulation
sidebands dominate the light in the interferometer output, i.e.\ the light
power of the sidebands must be larger than the 
light power of higher transverse modes leaving the interferometer
due to imperfections of the beam splitter and mirrors.

{\bf Detuned recycling} demands a different 
resonance condition in the signal recycling cavity.
Often only one signal
sideband is resonantly enhanced, whereas the gain for the other sideband can be
very low. 
Hence for greater detunings one signal sideband can be neglected,  
and the maximum sensitivity of the detuned interferometer  
is half the sensitivity 
of the interferometer in the broadband case at DC (assuming the same 
reflectivity for $\rm M_{\rm SR}$).
Since the resonance frequency of the signal recycling cavity now 
differs from that
of the power recycling cavity, the Schnupp modulation sideband 
throughput is more complex; thus the simple two-cavity model 
is no longer valid. Therefore 
the computer model was used to find control signals 
for the operating point $\xi_{\rm MI}= 0$ of the Michelson 
and the new operating point $\xi_{\rm SR} \not= 0$ for the
signal recycling mirror, preferably by using only one Schnupp modulation
frequency.
\begin{figure}[thb]
\IG [scale=.6] {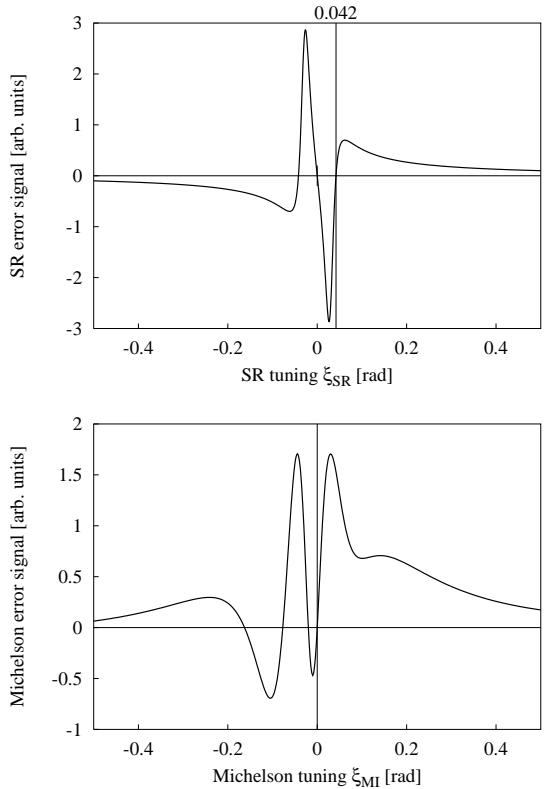} 
\caption{Simulated error signals for the signal recycling cavity and the
Michelson interferometer as a function of the signal recycling cavity 
length and the differential Michelson interferometer arm length. The 
conditions are those of the experiment shown in Figure~\ref{fig:mess}.}
\label{fig:err}
\end{figure}

The basic idea is to use the same photo detectors (i.e.~PD1~and~PD2)  but to 
shift the Schnupp modulation frequency by an offset of approximately  
$f_{\rm det}$ (the frequency of the sensitivity maximum) 
compared with the broadband case (see Fig.\ \ref{fig:det}). For example,
if the Schnupp modulation frequency 
in the broadband case was $f_{\rm mod}=9.69\,{\rm MHz}$, the new Schnupp
modulation frequency for an expected signal frequency
of 100\,kHz  would be
$f_{\rm mod}=9.69\,{\rm MHz}+100\,{\rm kHz}=9.79\,{\rm MHz}$. 
(With the free spectral range of the signal recycling cavity 
of $4.85\,{\rm MHz}$ the detuning to $100\,{\rm kHz}$
corresponds to an operating point of  
$\xi_{\rm SR} = 0.065$).
Figure\ \ref{fig:err}
shows the computed error signals for such a setup. This modulation
frequency gives the steepest slope and widest locking range for the
signal recycling control but results in a narrow locking 
range for the Michelson. Reducing the frequency shift $f_{\rm det}$ improves
the Michelson locking range with a slightly lower discriminator slope for the
signal recycling mirror.

\begin{figure}[bht]
\IG [scale=.6] {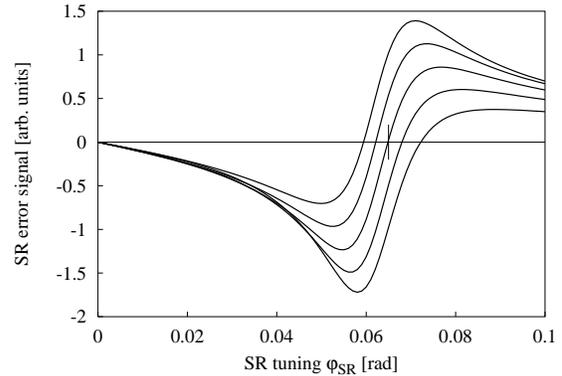} 
\caption{Simulated signal recycling error signals for different demodulation 
phases.}
\label{fig:chi}
\end{figure}
One feature of the discussed control scheme is that the operating point
of the signal recycling cavity is determined by both the Schnupp modulation
frequency {\em and\/} the demodulation phase $\chi_{\rm SR}$ that is used
to generate the signal recycling error signal. Figure\ \ref{fig:chi} shows
the dependence of the signal recycling error signal on $\chi_{\rm SR}$.
By varying $\chi_{\rm SR}$ by $\pm 0.5$\,rad the zero crossings move 
between 0.0594\,rad and 0.0722\,rad, which corresponds to sensitivity
peaks between 92\,kHz and 111\,kHz. Thus the rough tuning is 
determined by the shift of the 
modulation frequency, while fine tuning is possible by changing 
the demodulation phase 
$\chi_{\rm SR}$.
Furthermore, the operating point of the interferometer can be continously 
shifted from broadband mode to any desired detuning by simultaneous
adjustments of $f_{\rm mod}$,\ $\chi_{\rm SR}$ and $\chi_{\rm MI}$. 
This feature was used in our experiment.

\section{Experimental demonstration of detuned dual recycling}
In our experiment, the control signal needed to keep
the power recycling cavity resonant to the laser frequency was obtained 
independently with the standard Pound-Drever-Hall method~\cite{schnier:pr}.
An additional reference cavity was used for pre-stabilising the laser
frequency (see \cite{ghh:phd} for a detailed description of the experiment). 
To keep the setup as simple as possible it is desirable to use
only one Schnupp modulation frequency to generate both remaining
longitudinal control signals for the Michelson interferometer
and the signal recycling mirror. 
By demodulating the photocurrent 
of the South port photodetector PD1 with the
Schnupp modulation frequency an output signal for controlling the
Michelson interferometer is obtained (see Fig.~\ref{fig:block1}).

Using the same Schnupp frequency but the photocurrent from 
PD2, which detects
a weak beam picked off from one interferometer arm, a control
signal for the signal recycling mirror is obtained.
The mixers used for demodulation are driven by a local oscillator at
the Schnupp modulation frequency with the phases $\chi_{\rm MI}$ and
$\chi_{\rm SR}$. For the 30\,m prototype a Schnupp modulation frequency of 
$f_{\rm mod}\approx 2 \,\, {\rm FSR_{SR}=9.69\, MHz}$
was used.

To demonstrate detuned dual recycling the 30\,m interferometer was first 
locked in {\em broadband\/} dual recycling condition. 
The demodulation phases of the local oscillators
$\chi_{\rm MI}$ and $\chi_{\rm SR}$ were adjusted for 
maximum response of the Michelson interferometer error signal
at PD1 to a differential arm length
test signal.

\begin{figure}[htb]
\IG [scale=.6] {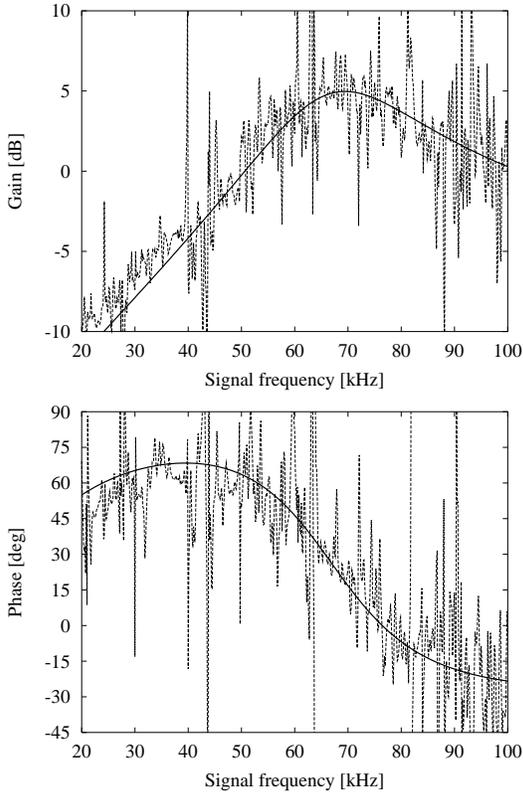} 
\caption{Measured transfer function and theoretical curve for detuned dual 
recycling.}
\label{fig:mess}
\end{figure}

A transfer function was measured using
coil-magnet actuators at the East end mirror to introduce a test signal of
known amplitude in the Michelson error signal (measured at photodetector
PD1). The ratio between the Michelson error
signal and the current through the actuator coils will be called the
{\em gain\/} of the interferometer.
Then the same actuators were used to apply a test signal of
69\,kHz, the next Fourier frequency at which the detuned dual recycling
was to be optimised.
In order to shift the operating point from broadband to detuned
the Schnupp modulation frequency and the demodulation 
phases $\chi_{\rm SR}, \chi_{\rm MI}$  were slowly and carefully adjusted for maximum gain, 
i.e.\ such that the
test signal appeared with maximal amplitude in the 
error signal of the Michelson interferometer. 
The best modulation frequency found was $9.736\,{\rm MHz}$. In this 
state the transfer function of the interferometer was measured. The 
difference in gain between the two measurements (i.e.\ the ratio of the two 
transfer functions) was expected to show the effect of the detuning 
while cancelling other effects like mechanical mirror resonances. 
Figure\ \ref{fig:mess}
shows the first results obtained. The transfer function shows the 
expected additional gain of about 5\,dB at around 69\,kHz. This 
means that after changing from broadband operation to detuned dual recycling 
the (test) signal at 69\,kHz
was detected in the main interferometer output 
with a 5\,dB higher amplitude. 

The theoretical curve shown in Fig.\ \ref{fig:mess} was obtained by 
calculating two transfer functions separately for the conditions
of the respective measurements. In the simulation the Schnupp 
modulation frequency was
changed from $9.688$ to $9.736\,{\rm MHz}$, 
the detuning of the signal recycling mirror
($\xi_{\rm SR}$) from 0.0 to 0.042\,rad, and the demodulation phase
of the Michelson ($\chi_{\rm MI}$) from -0.47 to -0.165\,rad.
The values for $\xi_{\rm SR}$ and $\chi_{\rm MI}$ had been determined
beforehand using a non-linear least square optimisation in which the
transfer function had been optimised for maximum gain at 69\,kHz by only
varying $\xi_{\rm SR}$ and $\chi_{\rm MI}$. 
This computation reflects the experimental procedure.

The good prediction of the 
measured additional gain around 69\,kHz and especially the good
agreement of the phase shift shows that the model is correct
and that the interferometer was indeed locked in a detuned state.

The locking range for the
Michelson can  be improved by changing the Schnupp modulation frequency
or the Michelson
demodulation phase from its optimum value at the expense of having 
a reduced Michelson 
interferometer discriminator slope
at $\xi_{\rm MI}=0$. 
Such readjustments can be used to increase the
locking range during lock acquisition.
Then the demodulation phases can be slowly tuned back to their optimal 
values with respect
to the gain (or discriminator slope) during operation.
In principle optimum values for locking range and gain 
can be achieved simultaneously by applying
separate modulation frequencies for the two control systems.

While the additional gain due to the detuning would be much larger in 
GEO\,600, 
the ratio between detuning and bandwidth is very similar in GEO\,600 
and our experiment.
The bandwidth of the 30\,m prototype interferometer, 
which is defined by the bandwidth of the signal
recycling cavity, was about 30\,kHz. A typical operation of the GEO\,600
detector would use a bandwidth of e.g.\ 200\,Hz and a detuning of around
500\,Hz (with a total detuning range of 0 to 1.5\,kHz), so that the ratio
between the bandwidth and detuning will be very similar to the one used here.

The detuned system behaved similar to the broadband case~\cite{ghh:dr}; 
given a good
initial alignment the lock acquisition happened by itself and the autoalignment
system worked as well as before. We also again observed the {\em mode healing\/} 
effect, which increases the contrast of the Michelson interferometer by
a factor of ten.

More measurements were carried out to test different operating points.
Using a fixed Schnupp modulation frequency of 9.740\,MHz, the detuning
$\xi_{\rm SR}$ was changed from 0.044 to 0.057\,rad, corresponding to
a shift of the sensitivity peak by 20\,kHz. 
Note that for locking the interferometer to different  detuned states only 
the signal recycling demodulation phase had to be re-adjusted. This 
might also be done by an electronic servo, which
would allow the adjustment of the sensitivity peak
of the detector automatically.

The experiment described is the first demonstration of detuned dual recycling 
with a fully suspended interferometer. The results are in good agreement 
with numerical simulations. Detuned dual recycling showed the same
operational advantages as broadband dual recycling. With the developed 
control scheme
which is fully compatible with the control of the broadband setup,
an automated tracking and stabilising of the peak sensitivity of GEO\,600
seems possible.

\ack
KAS and KDS would like to acknowledge support by PPARC and the 
University of Glasgow. KDS is also supported by a BP/RSE fellowship.
The authors are grateful for help and assistance from the GEO\,600 team.

\end{document}